\documentclass[11pt]{article}
\begin{document}

\title{ 
{\bf Supersymmetry in Singular Quantum Mechanics}}
\author{Ashok Das \\
\\
Department of Physics and Astronomy, \\
University of Rochester,\\
Rochester, New York, 14627}
\date{}
\maketitle

\begin{abstract}
The breaking of supersymmetry due to singular potentials
in supersymmetric quantum mechanics is critically analyzed. It
is shown that, when properly regularized, these potentials
respect supersymmetry, even when the regularization parameter is removed. 
\end{abstract}

\section{Introduction}

Supersymmetry is a beautiful and, simultaneously, a tantalizing
symmetry [1-7]. On the one hand, supersymmetry leads to field theories and
string theories with exceptional properties [8-9]. On the other hand,
supersymmetry also predicts degenerate superpartner states which are
not  observed experimentally and, consequently, one expects
that supersymmetry must be spontaneously (dynamically)
broken. However, unlike ordinary symmetries, spontaneous breaking of
supersymmetry has so far proved extremely difficult in the
conventional framework. Consequently, in the context of supersymmetry,
one constantly looks for alternate, unconventional methods of breaking
of this symmetry [6-7]. There is, of course, the breaking of supersymmetry
due to instanton effects which is well understood. However, several
authors, in recent years have suggested that supersymmetry may be
broken in the presence of singular potentials or boundaries in a
nonstandard manner [10-12]. 
The examples, where such a breaking has been discussed, are simple
quantum mechanical models which nonetheless arise from the
non-relativistic limit of some field theories. It is for this reason
that, in an earlier paper, we had examined [13] a candidate relativistic
$2+1$ dimensional field theory to see if the manifestation of such a
mechanism was possible in a field theory. However, a careful
examination  of the theory revealed that
supersymmetry prevails at the end although  it might appear naively, in
the beginning, that  supersymmetry would be broken in the nonstandard
manner. This prompted us to re-analyze the quantum mechanical models,
where this mechanism was demonstrated, more carefully and a
systematic and  critical examination,
once again, reveals [14] that  supersymmetry is
manifest even in such singular quantum mechanical models which is the
main result of this talk.

Since our discussion would be entirely within the context of one
dimensional supersymmetric quantum mechanics, let us establish the
essential notations here. Given a superpotential, $W(x)$, we can
define a pair of supersymmetric potentials as
\begin{equation}
 V_{+} = {1\over 2}\left(W^{2}(x) + W'(x)\right),\qquad V_{-} =
 {1\over 2}\left(W^{2}(x) - W'(x)\right)\label{a1}
\end{equation}
where \lq\lq prime'' denotes differentiation with respect to $x$. With
$\hbar =1$ and $m=1$, we can, then, define a pair of Hamiltonians
which describe a supersymmetric system as
\begin{equation}
H_{+} = - {1\over 2}\frac{d^{2}}{dx^{2}} + V_{+},\quad
H_{-} = - {1\over 2}\frac{d^{2}}{dx^{2}} + V_{-}\label{a2}
\end{equation}
In fact, defining the supercharges as
\begin{equation}
Q = {1\over\sqrt{2}}\left(-\frac{d}{dx} + W(x)\right), \qquad
Q^{\dagger} = {1\over\sqrt{2}}\left(\frac{d}{dx} + W(x)\right)\label{a3}
\end{equation}
we recognize that we can write the pair of Hamiltonians in
eq. (\ref{a2}) also as
\begin{equation}
H_{+} = Q^{\dagger}Q, \qquad H_{-} = QQ^{\dagger}\label{a4}
\end{equation}
All the
eigenstates of the two Hamiltonians $H_{+}$ and $H_{-}$ would be
degenerate except for the ground state with vanishing energy which
would correspond to the state satisfying
\begin{equation}
Q|\psi_{+}\rangle = 0, \qquad {\rm or,}\qquad Q^{\dagger}|\psi_{-}\rangle =
0\label{a6} 
\end{equation}

For a given superpotential, at  most one of the two conditions in
eq. (\ref{a6}) can be satisfied (that is, at  most, only one of the
two conditions in (\ref{a6}) would give a normalizable state). Namely,
the ground state with vanishing energy is unpaired and can belong to
the spectrum of either $H_{+}$ or $H_{-}$ depending on which of the
conditions leads to a normalizable state. This corresponds to the case
of unbroken supersymmetry. 
If, on the other hand, the superpotential is such that neither of the
states  in eq. (\ref{a6}) is normalizable, then, supersymmetry is
known  to be broken by instanton effects [6]. 
In this talk, we carefully analyze the models [10-12] where
supersymmetry is thought to be broken because of singular nature of
the potentials and show that when carefully analyzed, the systems with
singular potentials have manifest supersymmetry.

\section{Super \lq\lq Half'' Oscillator}

Let us consider a particle moving in the harmonic oscillator potential
on the \lq\lq half'' line
\begin{equation}
V(x) = \left\{\begin{array}{lll}
               {1\over 2}(\omega^{2}x^{2}-\omega)& {\rm for}& x>0\\
                \infty & {\rm for} & x<0
              \end{array}\right.\label{b1}
\end{equation}
The spectrum of this potential is quite clear intuitively. Namely,
because of the infinite barrier in the negative axis, we expect the
wave function to vanish at the
origin leading to the conclusion that, of all the solutions of the
oscillator on the full line, only the odd solutions (of course, on
the \lq\lq half'' line there is no notion of even and odd) would survive in
this case. While this is quite obvious, let us analyze the problem
systematically for later purpose.

First, let us note that singular potentials are best studied in a
regularized manner because this is the only way that  appropriate
boundary conditions can be determined correctly. Therefore, let us
consider the particle moving in the
regularized potential
\begin{equation}
V(x) = \left\{\begin{array}{lll}
               {1\over 2}(\omega^{2}x^{2}-\omega)& {\rm for}& x>0\\
                  &   &\\
               {c^{2}\over 2} & {\rm for} & x<0
              \end{array}\right.\label{b2}
\end{equation}
with the understanding that the limit $|c|\rightarrow\infty$ is to be
taken at the end. The Schr\"{o}dinger equation
can now be solved in the two regions. Since $|c|\rightarrow\infty$ at
the end, for any finite energy solution, we have the asymptotically
damped solution, for $x<0$,
\begin{equation}
\psi^{(II)}(x) = A\,e^{(c^{2} - 2\epsilon)^{{1\over 2}}\,x}\label{b3}
\end{equation}

Since the system no longer has reflection symmetry, the solutions, in
the region $x>0$,
cannot be classified into even and odd solutions. Rather, the
normalizable (physical)
solution would correspond to one which vanishes asymptotically. The
solutions of the Schr\"{o}dinger equation, in the region $x>0$, are
known as the parabolic cylinder functions [15] and the asymptotically
damped physical solution is given by
\begin{equation}
\psi^{(I)}(x) = B\,U(-({\epsilon\over\omega}+{1\over 2}),
\sqrt{2\omega}\,x)\label{b4} 
\end{equation}
It is now straightforward to match the solutions in eqs. (\ref{b3},
\ref{b4}) and their first derivatives across the boundary at $x=0$ and
their ratio gives
\begin{equation}
\frac{1}{\sqrt{c^{2}-2\epsilon}} =
-\frac{1}{2\sqrt{\omega}}\frac{\Gamma(-{\epsilon\over
2\omega})}{\Gamma(-{\epsilon\over 2\omega}+{1\over 2})}\label{b6}
\end{equation}
It is clear, then, that as $|c|\rightarrow \infty$, this can be
satisfied only if
\begin{equation}
-{\epsilon\over 2\omega} + {1\over 2}\;
 \stackrel{|c|\rightarrow\infty}{\longrightarrow}\; -n,\qquad\qquad
 n=0,1,2,\cdots\label{b7} 
\end{equation}
In other words, when the regularization is removed, the energy levels
that survive are the odd ones, namely, (remember that the zero point
energy is already subtracted out in (\ref{b1}) or (\ref{b2}))
$\epsilon_{n} = \omega (2n + 1)$.
The corresponding physical wave functions are nontrivial only on the half
line $x>0$ and have the form
\begin{equation}
\psi_{n}(x) = B_{n}\,U(-(2n+{3\over 2}), \sqrt{2\omega}\,x) =
\tilde{B}_{n}\,e^{-{1\over 2}\omega
x^{2}}\,H_{2n+1}(\sqrt{\omega}\,x)\label{b9}
\end{equation}
Namely, only the odd Hermite polynomials survive leading to the fact
that the wave function vanishes at $x=0$. Thus, we see that the
correct boundary condition naturally arises from regularizing the
singular potential and studying the problem systematically.

We now turn to the analysis of the supersymmetric oscillator on
the half line. One can define a superpotential [10]
\begin{equation}
W(x) = \left\{\begin{array}{cll}
              -\omega x & {\rm for} & x>0\\
              \infty & {\rm for} & x<0
              \end{array}\right.\label{b9'}
\end{equation}
which would, naively, lead to the pair of potentials
\begin{equation}
V_{\pm}(x) = \left\{\begin{array}{cll}
                    {1\over 2}(\omega^{2} x^{2} \mp \omega) & {\rm
                    for} & x>0\\
                    \infty & {\rm for} & x<0
                    \end{array}\right.
\end{equation}
Since, this involves  singular potentials, we can study it, as before,
by  regularizing the singular potentials as 
\begin{eqnarray}
V_{+}(x) & = &  \left\{\begin{array}{lll}
               {1\over 2}(\omega^{2}x^{2}-\omega)& {\rm for}& x>0\\
                  &   &\\
               {c_{+}^{2}\over 2} & {\rm for} & x<0
              \end{array}\right.\nonumber\\
V_{-}(x) & = &  \left\{\begin{array}{lll}
               {1\over 2}(\omega^{2}x^{2}+\omega)& {\rm for}& x>0\\
                  &   &\\
               {c_{-}^{2}\over 2} & {\rm for} & x<0
              \end{array}\right.\label{b10}
\end{eqnarray}
with the understanding that $|c_{\pm}|\rightarrow\infty$ at the end.

The earlier analysis can now be repeated for the pair of potentials in
eq. (\ref{b10}). It
is straightforward and without going into details, let us simply note
the results, namely, that, in this case, we obtain
\begin{eqnarray}
\epsilon_{+,n} & = & \omega(2n+1)\qquad
\psi_{+,n}(x)=B_{+,n}\,e^{-{1\over 2}\omega
x^{2}}\,H_{2n+1}(\sqrt{\omega}\,x)\nonumber\\
\epsilon_{-,n} & = & 2\omega(n+1)\qquad
\psi_{-,n}(x)=B_{-,n}\,e^{-{1\over 2}\omega
x^{2}}\,H_{2n+1}(\sqrt{\omega}\,x)\label{b11}
\end{eqnarray}
Here $n=0,1,2,\cdots$. There are several things to note from this
analysis. First, only the odd Hermite polynomials survive as physical
solutions since the wave function has to vanish at the origin. This
boundary condition
arises from a systematic study involving a regularized
potential. Second, the energy levels for the supersymmetric pair of
Hamiltonians are no longer degenerate. Furthermore, the state with
$\epsilon = 0$ no longer belongs to the Hilbert space (since it
corresponds to an even Hermite polynomial solution). This leads to the
conventional conclusion that supersymmetry is broken in such a case
and let us note, in particular, that in such a case, it would appear
that the superpartner states do not belong to the physical Hilbert
space (Namely, in this case, the supercharge is an odd operator and
hence connects even and odd Hermite polynomials. However, the boundary
condition selects out only odd Hermite polynomials as belonging to the
physical Hilbert space.).

There is absolutely no doubt that supersymmetry is broken in this
case. The question that needs to be addressed is whether it is a
dynamical property of the system or an artifact of the regularization
(and, hence the boundary condition) used. The answer is quite obvious,
namely, that
supersymmetry is broken mainly because the regularization (and,
therefore, the boundary condition) breaks
supersymmetry. In other words, for any value of the regularizing
parameters, $c_{\pm}$ (even if $|c_{+}|=|c_{-}|$), the pair of
potentials in eq. (\ref{b10}) do not define a supersymmetric system
and hence the regularization itself breaks
supersymmetry. Consequently, the breaking of supersymmetry that
results when the regularization is removed cannot be trusted as a
dynamical effect.

\subsection*{Regularized Superpotential}

Another way to understand this is to note that for a supersymmetric system,
it is not the potential that is fundamental. Rather, it is the
superpotential which gives the pair of supersymmetric potentials
through  Riccati type relations. It is natural, therefore, to
regularize the superpotential which would automatically lead to a pair of
regularized potentials which would be supersymmetric for any value of
the regularization parameter. Namely, such a regularization will
respect supersymmetry and, with such a regularization, it is, then, 
meaningful to ask if supersymmetry is broken when the regularization
parameter is removed at the end. With this in mind, let us look at the
regularized superpotential
\begin{equation}
W(x) = -\omega x\theta(x) + c\theta(-x)\label{b12}
\end{equation}
Here $c$ is the regularization parameter and we are supposed to take
$|c|\rightarrow\infty$ at the end. Note that the existence
of a normalizable ground state, namely, the form of the superpotential
in eq. (\ref{b9'}) selects out $c>0$
(otherwise, the regularization would have broken supersymmetry through
instanton effects as we have mentioned earlier).

The regularized superpotential now leads to the pair of regularized
supersymmetric potentials
\begin{eqnarray}
V_{+}(x) & = & {1\over 2}\left[(\omega^{2}x^{2}-\omega)\theta(x) +
c^{2}\theta(-x) - c\delta(x)\right]\nonumber\\
V_{-}(x) & = & {1\over 2}\left[(\omega^{2}x^{2}+\omega)\theta(x) +
c^{2}\theta(-x) + c\delta(x)\right]\label{b13}
\end{eqnarray}
which are supersymmetric for any $c>0$. Let us note that the
difference here from the earlier case where the potentials were
directly regularized (see eq. (\ref{b10})) lies only in the presence
of the $\delta(x)$ 
terms in the potentials. Consequently, the earlier solutions in the
regions $x>0$ and $x<0$ continue to hold. However, the matching
conditions are now different because of the delta function terms. 
Carefully matching the wave function and the discontinuity of the
first derivative across $x=0$ for each of the wavefunctions and taking
their ratio, we obtain the two conditions
\begin{eqnarray}
\frac{1}{(c^{2}-2\epsilon_{+})^{1/2} - c} & = & -{1\over
2\sqrt{\omega}}\frac{\Gamma(-{\epsilon_{+}\over
2\omega})}{\Gamma(-{\epsilon_{+}\over 2\omega}+{1\over
2})}\label{b14}\\
  &  & \nonumber\\
\frac{1}{(c^{2}-2\epsilon_{-})^{1/2} + c} & = & -{1\over
2\sqrt{\omega}}\frac{\Gamma(-{\epsilon_{-}\over
2\omega}+{1\over 2})}{\Gamma(-{\epsilon_{-}\over
2\omega}+1)}\label{b15}
\end{eqnarray}
It is now clear that, as $c\rightarrow\infty$, (\ref{b14}) and
(\ref{b15}) give respectively,
$\epsilon_{+,n} = 2\omega n$ and
$\epsilon_{-,n} = 2\omega(n+1)$ with $n=0,1,2,\cdots$.
The corresponding wave functions, in this case, have the forms
\begin{eqnarray}
\psi_{+,n}(x) & = & B_{+,n}\,e^{-{1\over 2}\omega
x^{2}}\,H_{2n}(\sqrt{\omega}\,x)\nonumber\\
\psi_{-,n}(x) & = & B_{-,n}\,e^{-{1\over 2}\omega
x^{2}}\,H_{2n+1}(\sqrt{\omega}\,x)\label{b17}
\end{eqnarray}
This is indeed quite interesting for it shows that the spectrum of
$H_{+}$ contains the ground state with vanishing energy. Furthermore,
all the other states of $H_{+}$ and $H_{-}$ are degenerate in energy
corresponding to even and odd Hermite polynomials as one would expect
from superpartner states. Consequently, it is quite clear that if the
supersymmetric \lq\lq half'' oscillator is defined carefully by
regularizing the superpotential, then, supersymmetry is manifest in
the limit of removing the regularization. This should be  contrasted
with  the general belief that supersymmetry is broken in this system
(which is a consequence of using boundary conditions or, equivalently,
of regularizing the potentials in a manner which violates
supersymmetry).
Of course, we should worry at this point as to how regularization
independent our conclusion really is. Namely, our results appear to
follow from the matching conditions in the presence of singular delta
potential terms and, consequently, it is worth investigating whether
our conclusions would continue to hold with an alternate
regularization of the superpotential which would not introduce such
singular terms to the potentials. We have done this [14] which shows that
our result is regularization independent.

\section{Oscillator with ${1\over x^{2}}$ Potential}

In the last section, we showed that, in the presence of one kind of
singularity, namely, a boundary, supersymmetry is unbroken. In what
follows, we will study another class of supersymmetric models, namely,
the supersymmetric oscillator with a ${1\over x^{2}}$ potential, where
there is a genuine singularity in the potential not necessarily
arising  from a boundary. A naive analysis of this model [11] also
shows that supersymmetry is broken by such a singular potential (for
certain parameter ranges). However, this conclusion can be understood,
again, as a consequence of regularizing the potential
which, as we have seen before, does not respect supersymmetry. In
stead, we will show through a careful analysis that,
when the superpotential is regularized, supersymmetry is manifest in
this model as well (with a lot of interesting features).  
In this section, however, we will systematically analyze only the quantum
mechanical system corresponding to an oscillator in the presence of a
${1\over x^{2}}$ potential (postponing the discussion of the
supersymmetric  case to the next section). This system has been
analyzed  by several people [16-18]  and
the most complete analysis appears to be in ref. [18]. However, we feel
that, while the energy levels derived in [18] are correct, the wave
functions are not (namely, the extensions of the solutions from the
positive to the negative axis are incomplete and the wave functions,
of course, become quite crucial  when one
wants to extend the analysis to a supersymmetric system) and,
consequently, we  present a careful analysis of this system
regularizing the singular potential in a systematic manner. With the
supersymmetric system in mind (to follow in the next section), we
write the potential for the system as (with $\hbar=m=\omega=1$)
\begin{equation}
V(x) = {1\over 2}\left[\frac{g(g+1)}{x^{2}} + x^{2} - 2g +
1\right]\label{c1}
\end{equation}

The singular potential is repulsive for $g>0$ or $g<-1$ while it is
attractive for $-1<g<0$. It is also worth noting here that the
Schr\"{o}dinger equation, in this case, is invariant under
\begin{eqnarray}
g & \leftrightarrow & -(g+1)\nonumber\\
\epsilon & \leftrightarrow & \epsilon + 2g + 1\label{c4}
\end{eqnarray}
This symmetry, of course, would also be reflected in the solutions.
Furthermore, 
the fixed point of this symmetry, namely, $g=-{1\over 2}$ separates
the two branches (namely, for every value of $\lambda$ there exist two
distinct values of $g$ corresponding to two distinct branches
separated at the branch point) in the parameter space.

\subsection*{Regularized Potential}

The Schr\"{o}dinger equation  can be solved quite easily
for $x>0$ as was also done in [18]. However, to determine correctly
how this wavefunction should be extended to the negative axis, it is
more suitable to regularize the potential near the origin and study
the problem carefully. Let us consider a potential of the form
\begin{equation}
V(x) = \left\{\begin{array}{lll}
{1\over 2}\left[\frac{g(g+1)}{x^{2}}+x^{2}-2g+1\right] & {\rm for} &
|x|>R\\
 & & \\
{1\over 2}\left[\frac{g(g+1)}{R^{2}}+R^{2}-2g+1\right] & {\rm for} &
|x|<R
\end{array}\right.\label{c5}
\end{equation}
Namely, we have regularized the potential in a continuous manner
preserving the symmetry in eq. (\ref{c4}) with the understanding that
the regularization parameter $R\rightarrow 0$ at the end. With this
regularization, the Schr\"{o}dinger equation has to be analyzed in
three distinct regions. However, since the potential has reflection
symmetry, we need to analyze the solutions only in the regions
$-R<x<R$ and $x>R$.

The potential is a constant in the region $-R<x<R$ and hence the
Schr\"{o}dinger equation is quite simple here. The solutions can be
classified into even and odd ones and take the forms
\begin{equation}
\psi^{(II)even}(x) = A(R) \cosh \kappa x,\quad
\psi^{(II)odd}(x)  = B(R) \sinh \kappa x\label{c6}
\end{equation}
where we have defined
\begin{equation}
\kappa = \sqrt{{g(g+1)\over R^{2}}+R^{2}-(2\epsilon + 2g - 1)}\approx
{\sqrt{g(g+1)}\over R}\label{c7}
\end{equation}
Since $R$ is small (and we are to take the vanishing limit at the
end), the last equality holds only if $g\neq 0\,{\rm or}\, -1$ which we will
assume. The special values of $g$ corresponding to the absence of a
singular potential have to be treated separately and we will come back
to this at the end of this section. We note here that the
normalization constants, $A$ and $B$, can, in principle depend on the
regularization parameter which we have allowed for in writing down the
form of the solutions in eq. (\ref{c6}).

The potential is much more complicated in the region $x>R$. However,
the physical solution can be obtained in terms of confluent
hypergeometric functions [15] in the form, (for $x>0$) 
\begin{eqnarray}
\psi^{(I)}(x) & = & C(R)\,e^{-{1\over
2}x^{2}}\left[\frac{\Gamma(-g-{1\over 2})}{\Gamma({1\over
2}-g-{\epsilon\over 2})}x^{g+1}\,M(1-{\epsilon\over 2},g+{3\over
2},x^{2})\right.\nonumber\\
 &  &+\left. \frac{\Gamma(g+{1\over 2})}{\Gamma(1-{\epsilon\over
2})}x^{-g}\,M({1\over 2}-g-{\epsilon\over 2},-g+{1\over
2},x^{2})\right]\label{c14}
\end{eqnarray}
Once again, we have allowed for a dependence of the normalization
constant, $C$, on the regularization parameter, $R$. However, for a
nontrivial solution to exist, we require that
\[
C(R)\stackrel{R\rightarrow 0}{\longrightarrow} C\neq 0
\]

So far, we have the general solutions, in the two regions, where
energy  is not quantized and 
which should arise from the matching conditions. Furthermore, we have
not bothered to evaluate the solution in the region $x<-R$ which
clearly would be the same as in the region $x>R$. However, the
matching conditions would determine  how we should extend the solutions
in the region $x>R$ to the region $x<-R$. Therefore, let us now
examine the matching conditions systematically since there are two
possible  cases.

\noindent$(i)$ {\bf\underline{Even Solution}}

We can match the even solution of the region $-R<x<R$ and its
derivative with those of the region $x>R$ at $x=R$. Taking the ratio
and remembering that $R$ is small (which is to be taken to zero at the
end), we obtain to the leading order in $R$
\begin{equation}
\sqrt{g(g+1)}\tanh \sqrt{g(g+1)} =\frac{(g+1){\Gamma(-g-{1\over
2})\over \Gamma({1\over 2}-g-{\epsilon\over
2})}R^{g+1}-g{\Gamma(g+{1\over 2})\over \Gamma(1-{\epsilon\over
2})}R^{-g}}{{\Gamma(-g-{1\over
2})\over \Gamma({1\over 2}-g-{\epsilon\over
2})}R^{g+1}+{\Gamma(g+{1\over 2})\over \Gamma(1-{\epsilon\over
2})}R^{-g}}\label{c15}
\end{equation}
Since the left hand side is independent of $R$, for consistency, the
right hand side must also be and this can happen in two different
ways.

First, for $g>-{1\over 2}$, it is clear that relation (\ref{c15}) can
be satisfied if (we assume from now on that $n=0,1,2,\cdots$.)
\begin{equation}
 \epsilon_{n}  =  2(n+1) -
2f_{1}(g)R^{2g+1}\label{c16}
\end{equation}
with a suitable choice of $f_{1}(g)$.

On the other hand, for $g<-{1\over 2}$, if
\begin{equation}
 \epsilon_{n} =  (2n-2g+1) -
2f_{2}(g)R^{-2g-1}\label{c17}
\end{equation}
relation (\ref{c15}) can be satisfied with a suitable choice of
$f_{2}(g)$. It is clear that the two possible branches of the solution
simply reflect the symmetry in eq. (\ref{c4}).

This analysis shows that when the regularization is removed (namely,
$R\rightarrow 0$), we have an even extension of the solution of the
forms
$g>-{1\over 2}$, $\epsilon_{n} = 2(n+1)$
with
\begin{equation}
\psi_{n}(x) = C_{n}\,\frac{\Gamma(-g-{1\over 2})}{\Gamma(-g-{1\over
2}-n)}e^{-{1\over 2}x^{2}}\,M(-n,g+{3\over
2},x^{2})\left\{\begin{array}{cll}
                x^{g+1}& {\rm for} & x>0\\
                |x|^{g+1}& {\rm for} & x<0
               \end{array}\right.\label{c19}
\end{equation}
and
$g<-{1\over 2}$, $\epsilon_{n} = 2n -2g + 1$
with
\begin{equation}
\psi_{n}(x) = C_{n}\,\frac{\Gamma(g+{1\over 2})}{\Gamma(g+{1\over
2}-n)}e^{-{1\over 2}x^{2}}\,M(-n,-g+{1\over
2},x^{2})\left\{\begin{array}{cll}
                x^{-g}& {\rm for} & x>0\\
                |x|^{-g}& {\rm for} & x<0
               \end{array}\right.\label{c21}
\end{equation}

\noindent $(ii)$ {\bf\underline{Odd Solution}}
In a similar manner, we can determine the odd solutions which have the
forms 
$g>-{1\over 2}$, $\epsilon_{n} = 2(n+1)$
with
\begin{equation}
\psi_{n}(x) = C_{n}\,\frac{\Gamma(-g-{1\over 2})}{\Gamma(-g-{1\over
2}-n)}e^{-{1\over 2}x^{2}}\,M(-n,g+{3\over
2},x^{2})\left\{\begin{array}{cll}
                x^{g+1}& {\rm for} & x>0\\
                -|x|^{g+1}& {\rm for} & x<0
               \end{array}\right.\label{c24}
\end{equation}
and
$g<-{1\over 2}$, $\epsilon_{n} = 2n -2g + 1$
with
\begin{equation}
\psi_{n}(x) = C_{n}\,\frac{\Gamma(g+{1\over 2})}{\Gamma(g+{1\over
2}-n)}e^{-{1\over 2}x^{2}}\,M(-n,-g+{1\over
2},x^{2})\left\{\begin{array}{cll}
                x^{-g}& {\rm for} & x>0\\
                -|x|^{-g}& {\rm for} & x<0
               \end{array}\right.\label{c26}
\end{equation}

\subsection*{Understanding  of the Result}

The conclusion following from this analysis, therefore, is that every
energy level of this system is doubly degenerate. Both even and odd
extensions of the solution are possible for every value of the energy
level. The energy levels which we have obtained are, of course,
identical to those obtained in [18]. The crucial difference is in the
structure of the wave functions, namely, that both even and odd
extensions of the solution are possible for every value of the
energy (Incidentally, the solutions we have obtained in terms of
confluent hypergeometric functions also coincide with generalized
Laguerre polynomials as was obtained in ref. [18].). It is crucial,
therefore, to ask if such a conclusion is
physically plausible. To understand this question, let us recapitulate
the results from a simple quantum mechanical model which is well
studied. Namely, let us look at a particle moving in a potential of
the form
\[
V(x) = \left\{\begin{array}{cll}
              \gamma \delta(x) & {\rm for} & |x|<a\\
              \infty & {\rm for} & |x|>a
              \end{array}\right.
\]
It is well known that the solutions of this system can be classified
into even and odd ones with energy levels ($\hbar=m=1$)
\[
E_{n}^{even} = \frac{n^{2}\pi^{2}}{2(a+{1\over
\gamma})^{2}},\quad
E_{n}^{odd} = \frac{n^{2}\pi^{2}}{2a^{2}}
\]
The even and the odd solutions, of course, have distinct energy values for
any finite strength of the delta potential. However, when
$\gamma\rightarrow\infty$, both the even and the odd solutions become
degenerate in energy. Namely, a delta potential with an infinite
strength leads to a double degeneracy of every energy level corresponding
to both even and odd solutions. The connection of this example with
the problem we are studying is intuitively clear. Namely, we can think
of
\[
{g(g+1)\over x^{2}} = \lim_{\eta\rightarrow 0}\,{g(g+1)\over
x^{2}+\eta^{2}} = \lim_{\eta\rightarrow 0}\left({\pi g(g+1)\over
\eta}\right)\,\left({1\over \pi}{\eta\over x^{2}+\eta^{2}}\right)
\]
It is clear that for $g\neq 0\,{\rm or}\,-1$, the singular ${1\over x^{2}}$
potential behaves like a delta potential with an infinite strength and
it is quite natural, therefore, that this system has both even and odd
solutions degenerate in energy.

It is also clear from this analysis that it is meaningless to take the
$g=0\,{\rm or}\,-1$ limit from the results obtained so far simply
because  the characters
of the two problems are quite different. As we have argued, for any
finite value of $g$ not coinciding with those special values, the
potential behaves, at the origin, like a delta potential of infinite
strength while for the special values, there is no such potential. The
two cases are related in a drastically discontinuous manner. As a
result, one cannot treat the ${g(g+1)\over x^{2}}$ as a perturbation
and obtain the full, correct solution simply because there is nothing
perturbative (small) about this potential for any \lq\lq nontrivial''
value of $g$. Another way of saying this is to re-emphasize what we
have already observed following eq. (\ref{c7}), namely, the character
of $\kappa$ and, therefore, the matching conditions change depending
on whether or not $g$ differs from the special values $0,-1$.

\section{Supersymmetric Oscillator with ${1\over x^{2}}$ Potential:}
The supersymmetric version of the
case studied is obtained from a superpotential of
the form [11]
\begin{equation}
W(x) = {g\over x} - x\label{d1}
\end{equation}
In this case, it is easily seen that a normalizable ground state
wavefunction exists only for $g>-{1\over 2}$.

The superpotential and, therefore, the potential is singular at the
origin. Thus, once again, the proper way to study the spectrum of such
a system is by regularizing the superpotential. We introduce the
regularized superpotential
\begin{equation}
W(x) = \theta(x-|R|)\left({g\over x} - x\right) +
\theta(|R|-x)\left({g\over R}-R\right){x\over R}\label{d2}
\end{equation}
Here, $R$ denotes the regularization parameter which is to be taken to
zero at the end. The regularized superpotential is continuous and the
resulting pair of potentials take the forms
\begin{eqnarray*}
V_{+} & = & {1\over 2}\left[\theta(x-|R|)\left({g(g-1)\over
x^{2}}+x^{2}-2g-1\right) + \theta(|R|-x)\left(\left({g\over
R^{2}}-1\right)^{2}x^{2}+\left({g\over
R^{2}}-1\right)\right)\right]\nonumber\\
V_{-} & = & {1\over 2}\left[\theta(x-|R|)\left({g(g+1)\over
x^{2}}+x^{2}-2g+1\right) + \theta(|R|-x)\left(\left({g\over
R^{2}}-1\right)^{2}x^{2} - \left({g\over
R^{2}}-1\right)\right)\right]\label{d3}
\end{eqnarray*}

The solutions of the pair of Hamiltonians can now be studied. Without
going into details [14], let us note that the solutions, in this case,
again turn out to be confluent hypergeometric functions. There are
several interesting features that arise in this case. For example, it
turns out that, in the limit $R\rightarrow 0$, $H_{+}$ has three sets
of normalizable solutions -- one even and two odd. The three
sets of normalizable solutions of $H_{-}$ also correspond to one even
and two sets of odd solutions. While one of the three sets of
solutions correspond to a supersymmetric system, there are additional
solutions which apparently have no relation to one another.

The proper understanding of the solutions comes really from
recognizing that, given a bosonic system, there is an arbitrariness in
supersymmetrizing the system. It is much like the arbitrariness of
whether a spin ${1\over 2}$ particle belongs to a supersymmetric
multiplet $(0,{1\over 2})$ or $({1\over 2},1)$. The different
solutions really correspond to different possible supersymmetrizations
and matching has to be done carefully. When analyzed carefully, it
turns out that supersymmetry is manifest in the system. In addition,
in this case, the problem can be solved algebraically because of a
special symmetry in the problem known as shape invariance. The
algebraic solution also coincides with the explicit solutions obtained.

\section{Conclusion}
In this talk, we have discussed systematically two classes of
supersymmetric quantum  mechanical models - one consisting of a
singular boundary and the other with a singular potential. We have
shown that, contrary to the conventional understanding [10-12], supersymmetry
is manifest in these systems. In particular, for a system with a
singular potential such as ${1\over x^{2}}$,  the
solution of the Schr\"{o}dinger equation leads to several distinct
solutions corresponding to distinct supersymmetrizations of the
system. Consequently, it becomes quite important to identify the
appropriate wavefunctions when supersymmetric properties are being
investigated. Finally, we would like to conclude by noting that
supersymmetry is known to be robust at short distances (high
energies). The singularities discussed in the quantum mechanical
models occur at short distances  and, therefore, it is intuitively quite
clear that they are unlikely to break supersymmetry. Our detailed,
systematic analysis only reinforces this.    

\section*{Acknowledgments}

This work was supported in part by the U.S. Dept. of
Energy Grant  DE-FG 02-91ER40685 and NSF-INT-9602559.


\begin{thebibliography}{99}
\bibitem{GL} Y. A. Gol'fand and E. P. Likhtman, JETP Lett. {\bf 13}
(1971) 323.
\bibitem{R} P. Ramond, Phys. Rev. {\bf D3} (1971) 2415; A. Neveu and
J. Schwarz, Nucl. Phys. {\bf 31} (1971) 86.
\bibitem{VA} D. Volkov and V. Akulov, Phys. Lett. {\bf B46} (1973)
109.
\bibitem{WZ} J. Wess and B. Zumino, Nucl. Phys. {\bf 70} (1974) 39.
\bibitem{S} P. Fayet and S. Ferrara, Phys. Rep. {\bf 32} (1977)
249; M. F. Sohnius,  Phys. Rep. {\bf 128} (1985) 39.
\bibitem{W} E. Witten, Nucl. Phys. {\bf B188} (1981) 513; {\it ibid}
{\bf B202} (1982) 253; P. Salomonson and J. W. van Holten,
Nucl. Phys. {\bf B196} (1982) 509; F. Cooper and B. Freedman,
Ann. Phys. {\bf 146} (1983) 262.
\bibitem{BCD} C. Bender, F. Cooper and A. Das, Phys. Rev. {\bf D28}
(1983) 1473.
\bibitem{GGRS}S. J. Gates Jr, M. Grisaru, M. Roc\v{e}k, W. Siegel,
{\it Superspace or One Thousand and One Lessons in Supersymmetry}
(Benjamin/Cummings, Reading, Mass. 1983).
\bibitem{GSW} M. B. Green, J. Schwarz and E. Witten, {\it Superstring
Theory} (Cambridge, 1987).
\bibitem{RR} P. Roy and R. Roychoudhury,  Phys. Rev.
 {\bf D32} (1985), 1597.
\bibitem{jr} A. Jevicki, J. Rodrigues, Phys. Lett. {\bf B
146} (1984) 55.
\bibitem{cks} F. Cooper, A. Khare, and U. Sukhatme,
Phys. Rep.{\bf 251}, (1995) 267.
\bibitem{DP} A. Das and S. A. Pernice, Nucl. Phys. {\bf B505} (1997)
123.
\bibitem{DP2} A. Das and S. A. Pernice, Nucl. Phys. {\bf B561} (1999) 357.
\bibitem{ASt} M. Abramowitz and I. A. Stegun, {\it Handbook
of Mathematical Functions}, Dover Publications (1972).
\bibitem{Zi} F. Zirilli, J. Math. Phys. {\bf 15} (1974) 1202.
\bibitem{C} F. Cologero, J. Math. Phys. {\bf 10} (1969) 2191.
\bibitem{lat} L. Lathouwers, J. of Math. Phys., {\bf 16}
(1975) 1393.

\end{thebibliography}
\end{document}